\documentclass[reprint,notitlepage,aps,prd,showpacs,onecolumn]{revtex4-1}
\usepackage[utf8x]{inputenc}
\usepackage{amsmath, amssymb}

\begin{document}

\title{Uniqueness of the Fock quantization of scalar fields in spatially flat cosmological spacetimes}

\author{Laura Castell\'o Gomar}
\email{laucaste@estumail.ucm.es}
\affiliation{Facultad de Ciencias Físicas, Universidad Complutense de Madrid, Ciudad Universitaria, 28040 Madrid, Spain}

\author{Jer\'onimo Cortez}
\email{jacq@ciencias unam.mx}
\affiliation{Departamento de F\'{\i}sica, Facultad de Ciencias, Universidad Nacional Aut\'onoma de M\'exico, Mexico D.F. 04510, Mexico}

\author{Daniel \surname{Mart\'{\i}n-de~Blas}}
\email{daniel.martin@iem.cfmac.csic.es}

\author{Guillermo A. Mena~Marug\'an}
\email{mena@iem.cfmac.csic.es}
\affiliation{Instituto de Estructura de la Materia, CSIC, Serrano 121, 28006 Madrid, Spain}

\author{Jos\'e M. Velhinho}
\email{jvelhi@ubi.pt}
\affiliation{Dept. de F\'{\i}sica, Universidade da Beira Interior, R. Marqu\^es D'\'Avila e Bolama, 6201-001 Covilh\~a, Portugal.}

\begin{abstract}
We study the Fock quantization of scalar fields in (generically) time dependent scenarios, focusing on the case in which the field propagation occurs in --either a background or effective-- spacetime with spatial sections of flat compact topology. The discussion finds important applications in cosmology, like e.g. in the description of test Klein-Gordon fields and scalar perturbations in Friedmann-Robertson-Walker spacetime in the observationally favored flat case. Two types of ambiguities in the quantization are analyzed. First, the infinite ambiguity existing in the choice of a Fock representation for the canonical commutation relations, understandable as the freedom in the choice of inequivalent vacua for a given field. Besides, in cosmological situations, it is customary to scale the fields by time dependent functions, which absorb part of the evolution arising from the spacetime, which is treated classically. This leads to an additional ambiguity, this time in the choice of a canonical pair of field 
variables. We show that both types of ambiguities are removed by the requirements of (a) invariance of the vacuum under the symmetries of the three-torus, \emph{and} (b) unitary implementation of the dynamics in the quantum theory. In this way, one arrives at a unique class of unitarily equivalent Fock quantizations for the system. This result provides considerable robustness to the quantum predictions and renders meaningful the confrontation with observation.
\end{abstract}


\maketitle

\section{Introduction}
\label{s1}

The description of fields in nonstationary scenarios like those encountered in cosmology is plagued with severe ambiguities which undermine the robustness of the predictions up to the verge of becoming devoid of physical significance. Obviously, the existence of ambiguities in the construction of a quantum description for a classical system is an issue which affects all fields of physics. However, the problem is especially relevant in the case of cosmology because, on the one hand, the windows available for quantum phenomena are narrow (if any) and, on the other hand, one cannot carry out an unlimited number of repeated measurements on the system (or on a large number of copies of it). Rather, one observes the only available system, namely the Universe in which we live.

In general, the assignation of a quantum theory to a classical system is a process that involves many ambiguous choices, which can affect the physical predictions. For instance, one has a vast freedom in the possible selection of fundamental variables to describe the system, and in the algebra of observables constructed out of them which one wants to represent by quantum operators. Moreover, even if this basic step is fixed, one still finds representations that are inequivalent, and therefore lead to different physics in the quantum realm. For systems with finitely many degrees of freedom and whose phase space is linear, the freedom in the choice of a specific representation can be removed if one demands that the representation possess certain good properties, namely, unitarity, irreducibility, and strong continuity: the uniqueness of the representation is then guaranteed by the so-called Stone-von Neumann theorem \cite{simon}. Nevertheless, when one deals with systems with an infinite number of degrees of 
freedom, as in quantum field theory, the situation gets much more involved. Even in the simplest fieldlike systems and restricting the discussion to Fock representations, obtained via the action of creator operators on a vacuum state, there exist an infinite number of possible choices with an equally acceptable behavior from the mathematical viewpoint \cite{wald}. Therefore, unless one includes additional criteria \cite{wald,criteria,jackiw} to select a vacuum state (or rather a unitarily equivalent class of them), the Fock representation of the canonical commutation relations (CCR's) of the field has to face a fundamental, infinite ambiguity.

Moreover, in cosmological scenarios like those that we are interested to study, there exists an ambiguity which is previous to the choice of Fock representation, and which has to do in part with the commented freedom in the selection of a canonical pair of variables. This ambiguity arises in a natural way in systems defined in nonstationary settings, because in those frameworks it is customary to scale the field configurations by time varying functions which absorb part of the time dependence of the spacetime in which the field propagates. This is frequently the case irrespective of whether the spacetime is a true physical background \cite{birrell,mukhanov}, an effective spacetime (for instance, for fields which propagate in quantum corrected backgrounds, like in effective Loop Quantum Cosmology \cite{LQCap,fmov,effLQC,hybrid-pert}), or an auxiliary spacetime (e.g. in the case of dimensional reductions of gravitational systems with two commuting spacelike isometries, like the Gowdy models \cite{redGowdy,cm,
ccm}). Scalings of this type are found, for example, in the treatment of Klein-Gordon (KG) fields in Friedmann-Robertson-Walker (FRW) spacetimes, in order to transform the KG equation into a generalized wave equation in a static spacetime (as we will see below), or in the treatment of scalar perturbations around FRW spacetime, like in the introduction of Mukhanov-Sasaki variables \cite{muk-sas}. The scaling of the field configuration can be completed into a linear and time dependent canonical transformation. This transformation leads to a new canonical pair of field variables which, furthermore, has typically a different (but still linear) dynamics, since the changes performed are time varying.

From these considerations it is clear that, in order to extract significant quantum predictions for fieldlike systems in cosmology, or more generally in time dependent settings, one needs to introduce criteria which, based on reasonable physical motivations, can fix the commented ambiguities and pick out a unique class of quantum descriptions which are all unitarily equivalent (and hence lead all to the same physics).

A common procedure to select a unique Fock representation of the CCR's for a given field is to benefit from the classical symmetries of the system and demand that the vacuum be invariant under them. Notice that this requirement, though well founded inasmuch as it involves demanding that all the basic structures for quantization are symmetry invariant, it is more stringent than asking simply for a unitary representation of the group of symmetries in the quantum theory. For some highly symmetric backgrounds (as is the case of Minkowski spacetime), the invariance under the (Poincar\'e) group is enough to select a unique Fock quantization \cite{wald}. However, in general, when less symmetric backgrounds are considered, this criterion fails to select a unique representation of the CCR's by its own. In brief, there is not sufficient symmetry available and there still exist infinitely many inequivalent vacua which are invariant.

Nonetheless, this symmetry invariance remains a useful tool in order to select a preferred Fock representation when additional requirements are introduced. This is the case of cosmological systems, or in general systems whose linear field equations have no timelike symmetries. In these situations, it seems reasonable to replace the role of symmetry invariance with the weaker demand of unitarity in the time evolution. In other words, to complement the invariance under spatial symmetries with the requirement of a unitary dynamics of the system in its Fock description. This requirement is crucial unless one is willing to renounce to a standard probabilistic interpretation of the quantum theory, including the evolution of the corresponding observables. This criterion was proposed and applied for the first time in the literature when considering the quantization of gravitational waves in inhomogeneous Gowdy cosmologies \cite{cm,ccm,ccmv1,ccmv2,cmv,BVV2,CQG25}. These waves admit an alternate description in a 
dimensionally reduced auxiliary spacetime, in which they satisfy a field equation of KG type, with an effective mass which varies in time. The spatial sections of this auxiliary spacetime can be either the circle or the two-sphere, depending on the actual spatial topology of the Gowdy spacetime. Moreover, this criterion for uniqueness of the Fock quantization has been applied more recently to scalar fields with a(n almost) generic time dependent mass function and defined on any $d$-sphere, with $d$ less or equal to three \cite{PRD79,CMV8,CMOV-S3S}. Besides, the criterion has been slightly generalized (employing just the main part of the differential field equations) and used for the quantization of the perturbations of a minimally coupled massive scalar field in an FRW universe with positive curvature (with spatial sections isomorphic to three-spheres), as an example of an inflationary scenario \cite{fmov}. Finally, the selection of a unique Fock representation for KG fields in (auxiliary) static spacetimes 
in the presence of a(n effective) time dependent mass has been recently investigated in Ref. \cite{CMOV-FTC} for generic compact spatial sections in three or less dimensions. The generalization of the criterion to remove the ambiguity of the scaling of the field by background time dependent functions in the case of spatial sections with any kind of compact topology is under discussion \cite{CMOV-GUS}.

In this work, we will focus on the case of compact topology which is more relevant from the point of view of its application to cosmology: flat topology in three dimensions. The compactness of the topology is assumed to obviate possible infrared problems. Nonetheless, the effects of compactness should be ignorable if the physics does not depend on very large scales, e.g. those larger than a Hubble radius in cosmology. Specifically, we will consider the Fock quantization of a scalar field defined on a spatial manifold isomorphic to a three-torus and which, after possibly a suitable scaling by a time varying function, satisfies a KG equation with a time dependent mass. The relevance of the system in physical cosmology --as in the description of minimally coupled scalar fields and scalar perturbations in a flat FRW spacetime, which provides models that approximately describe the observed Universe-- together with a careful discussion of the proofs of our results for flat three-dimensional topology --emphasizing 
the peculiarities of this case and making the physics behind our line of arguments more accessible-- justifies a specialized analysis of this setting by its own.

Although we will present more details about the applications of our results in the concluding section, let us consider here as an example the simple case of a test scalar field  $\phi$, with mass $m$, that is minimally coupled to a flat FRW spacetime with sections isomorphic to a three-torus. The field satisfies the following equation (see e.g. \cite{full}):
\begin{equation}
\label{eqm-frw1}
\phi^{\prime\prime}+ 3\,\frac{a^{\prime}}{a}\,\phi^{\prime} -\frac{1}{a^{2}}\Delta \phi + m^2 \phi = 0,
\end{equation}
where $a$ denotes the (time dependent) scale factor of the FRW geometry, $\Delta$ is the Laplace-Beltrami (LB) operator for the standard (static) three-torus metric, and the  prime stands here for the derivative with respect to the proper time. Changing the field with the scale factor, $\varphi=a\phi$, and using conformal time, the field equation is transformed into the following one:
\begin{equation}
\label{eqm-frw2}
\ddot{\varphi}-\Delta \varphi +s\varphi = 0.
\end{equation}
Here the dot denotes the derivative with respect to the conformal time, and $s$ is a mass function given by $s=m^2 a^{2}-\ddot{a}/a$. From now on, we will consider systems that, via a convenient scaling, admit a field equation like Eq. \eqref{eqm-frw2}, allowing $s$ to be a(n  almost) generic time function.

As we have said, there exist certain peculiarities in the case of the three-torus topology which motivate the mathematical interest in its analysis. Among these peculiarities, let us mention that the group of isometries of the three-torus is an Abelian compact group (as in the $S^{1}$ case \cite{PRD79}). Therefore, the irreducible (unitary) representations are  one-dimensional and defined over complex vector spaces \cite{brocker}. This introduces some subtleties in the characterization of the structures on phase space that are invariant under the spatial symmetries, since complex representations must be combined so that, at the end of the day, one deals exclusively with real scalar fields. In particular, one must take this issue into account in the characterization of the invariant complex structures,  changing the approach followed for other topologies in the previous literature, like e.g. in the case of spheres \cite{CMV8,CMOV-S3S}.

Recall that a Fock representation of the CCR's is completely determined by the fixation of a complex structure \cite{wald}. A complex structure $J$ is a real linear map on phase space which provides a symplectic transformation --namely, it preserves the symplectic structure $\Omega$ of the system, which encodes the information about the Poisson brackets and the CCR's--  and whose square is minus the identity \cite{wald,ccq}. We further require the complex and the symplectic structures to be compatible, in the sense that one obtains a positive definite bilinear map when modifying $\Omega$ by first acting with $J$ on one of its entries. The election of the complex structure is equivalent to the election of annihilation and creationlike variables (up to physically irrelevant redefinitions), and hence to the choice of vacuum state for the quantum theory.

We are interested only in those complex structures that are invariant under the spatial symmetries of the system and lead to a unitary implementation of the field dynamics. In general, a linear canonical transformation $T$ can be implemented unitarily in the representation determined by a complex structure $J$ if and only if the antilinear part of the transformation [$T_J=(T+JTJ)/2$] is a Hilbert-Schmidt operator \cite{hr-sh} (i.e., $T_J^{\dagger}T_J$ has a finite trace, where the dagger denotes the adjoint operator). This condition amounts to demanding that the antilinear coefficients of the transformation (usually called the beta Bogoliubov coefficients) be square summable. In addition, this is equivalent to the condition that, under the action of $T$, the transformed vacuum contain a finite number of ``particles'' as seen by the original vacuum state.

We will show that the combined imposition of (a) the invariance of the complex structure (and therefore of the vacuum) under the isometries of the three-torus \emph{and} (b) the unitary implementation of the dynamics in the quantum theory, suffice to select a unique class of equivalent Fock representations for KG fields in static spacetimes with a time dependent mass. Actually, this class of representations contains the one associated with a free massless scalar field. Moreover, we will show that a Fock representation with the listed properties is possible only for fields that satisfy equations of the type \eqref{eqm-frw2}, a fact that completely removes the ambiguity in the choice of a canonical pair of fields among all those related by a time dependent scaling of the configuration, plus a time dependent linear redefinition of the momentum. This fixes the Fock quantization of the system, as far as the two kinds of ambiguities that we consider here are concerned.

The rest of the paper is organized as follows. In Sec. \ref{s2}, we introduce the field system and present a Fock quantization based on the complex structure which is naturally associated with a free massless scalar field. Then, in Sec. \ref{s3} we prove that the Fock representation determined by this structure implements the evolution as a unitary transformation (with a particular scaling of the field). We also show that any other Fock representation of the CCR's with unitary dynamics and invariant under the three-torus symmetries is unitarily equivalent. Sec. \ref{s4} deals with the ambiguity in the choice of a canonical pair for the system, among all those related by linear transformations which include a time dependent scaling of the field configuration. We show that, in order to obtain a unitarily implementable evolution, neither a field scaling nor a redefinition of the momentum is allowed. Finally, Sec. \ref{s5} contains the conclusions and a discussion of the results.

\section{The Klein-Gordon model}
\label{s2}

Let us start by considering a real scalar field $\varphi$ defined on a globally hyperbolic, static spacetime whose spatial sections are isomorphic to a three-torus, equipped with its standard spatial metric $h_{ij}$ ($i,j=1,2,3$). The time domain can be any arbitrary connected real interval  $\mathbb{I}\subset \mathbb{R}$. We do not impose $\mathbb{I}$ to be unbounded, nor of any specific form. This is important if we want our analysis to be applicable even in the case of effective spacetimes, which often can be treated classically (at least in certain approximations) only for some restricted time intervals (this might be the case, e.g., for effective descriptions derived from Quantum Cosmology \cite{LQCap}). In addition, we suppose that the scalar field is subject to an equation of KG type, with a quadratic potential of the form $V(\varphi)=s(t)\varphi^{2}/2$, which can be interpreted as a mass term varying in time. In principle, we allow $s(t)$ to be any time function. Later on, we will assume in our
discussion that it satisfies the mild condition of possessing a second derivative which is integrable in every compact subinterval of $\mathbb{I}$.

The canonical phase space $\Gamma$ of the field system is obtained from the Cauchy data $(\varphi, P_{\varphi})=(\varphi_{|t_{0}}, \sqrt{h}\dot{\varphi}_{|{t_{0}}})$ at a reference time $t_0$. The dot denotes time derivative, and we keep explicit the determinant $h$ of the spatial metric, even though it is equal to one for the three-torus, in order to facilitate the extension of our discussion to more general cases. This phase space is equipped with a symplectic structure $\Omega$ that amounts to the canonical Poisson brackets $\{\varphi(x),P_{\varphi}(y)\}=\delta(x-y)$, with $\delta(x)$ being the Dirac delta on the three-torus. Recall that the Poisson brackets, and therefore $\Omega$, are independent of the choice of ``initial'' time $t_{0}$.

The Hamilton equations of motion are
\begin{equation}
\label{fe1}
\dot{\varphi}=\frac{1}{\sqrt{h}}P_{\varphi}, \qquad \dot{P}_{\varphi}=\sqrt{h}[\Delta\varphi-s(t)\varphi],
\end{equation}
where $\Delta$ is the standard LB operator for the three-torus, completely determined by the (static) metric $h_{ij}$. It is clear that equations \eqref{fe1} imply equation \eqref{eqm-frw2}. It is important to recall again that this kind of equations can be obtained in very different cosmological models by considering a suitable canonical transformation which involves a time depending scaling of the field configurations \cite{CMOV-S3S} (and often the consideration of conformal time). In this way, the field equations can be interpreted as a KG equation with a time dependent mass in a static spacetime. On the other hand, note that these dynamical equations are invariant under the group of isometries of the three-torus. We can obtain such isometries from the composition of rigid rotations in each of the periodic spatial directions (diagonalizing the three-torus metric), $T_{\alpha_{i}}:\theta_{i}\rightarrow \theta_{i}+\alpha_{i}$ , with $\alpha_{i}\in S^{1}$, and $i=1,2,
3$. Let us introduce the notation $T_{\vec{\alpha}}$ for these transformations, where $\vec{\alpha}$ is the tuple $(\alpha_{1},\alpha_{2},\alpha_{3})$, and $T_{\vec{\alpha}}\equiv T_{\alpha_{1}}\circ T_{\alpha_{2}}\circ T_{\alpha_{3}}$.

We decompose the field (and its momentum) in a series expansion adapted to the three-torus symmetries using eigenfunctions of the LB operator. These eigenfunctions provide a basis in the space of square integrable functions on the three-torus with the volume element determined by $h_{ij}$. The series expansion can be done using complex eigenfunctions (namely, a Fourier basis of plane waves):
\begin{equation}
\label{cfd1}
\varphi(t,\vec{\theta})=\frac{1}{(2\pi)^{3/2}}\sum_{\vec{m}}\mathfrak{q}_{\vec
{m}}(t) \exp\{i(\vec{m}\cdot\vec{\theta})\},
\end{equation}
where $\vec{m}=(m_1,m_2,m_3)$, with $m_i\in \mathbb{Z}$, $i=1,2,3$, and $\vec{m}\cdot\vec{\theta}=\sum_{i=1}^3 m_i \theta_i$. Since the fields are real, one must impose the reality conditions
\begin{equation}
\mathfrak{q}_{-\vec{m}}(t)=\left(\mathfrak{q}_{\vec{m}}(t)\right)^{\ast
},
\end{equation}
where the symbol $\ast$ stands for complex conjugation. This complex decomposition is useful because each tuple of integer numbers $\vec{m}$ provides an irreducible representation of the isometry group. The drawback, on the contrary, is that one has to deal with the complications posed by the presence of reality conditions.

The field can be also decomposed employing real eigenfunctions (i.e., in a Fourier expansion in cosines and sines):
\begin{equation}
\label{rfd1}
\varphi(t,\vec{\theta})=\frac{1}{\pi^{3/2}}\sum_{j=0}^{3}\sum_{\vec{n}_{j}}\left[q_{\vec{n}_j}(t)\cos(\vec{n}_{j}\cdot\vec{\theta})
+x_{\vec{n}_j}(t)\sin(\vec{n}_{j}\cdot\vec{\theta})\right].
\end{equation}
Here, we have defined the tuple $\vec{n}_{0}=(n_1,n_2,n_3)$, with $n_i\in \mathbb{N}$, and the tuples $\vec{n}_{j}$ are obtained from $\vec{n}_{0}$ by changing the sign of $n_j$. It is important to stress that, when the tuple $\vec{n}_0$ contains one or more zeros, not all the modes $\{q_{\vec{n}_j}, x_{\vec{n}_j}\}$ are independent for the different possible values of $j$, but there are modes that are actually the same. We assume that they are considered only once in the above sum, without changing and complicating the introduced notation.

We decompose the field momentum in the very same way. Its real modes have coefficients satisfying $p_{\vec{n}_j}=\dot{q}_{\vec{n}_j}$, and $y_{\vec{n}_j}=\dot{x}_{\vec{n}_j}$ (for the cosine and sine contributions, respectively). In terms of the configuration and momentum Fourier coefficients, the nonvanishing Poisson brackets are $\{q_{\vec{n}_j},p_{\vec{n}_{j}'}\}=\{x_{\vec{n}_j},y_{\vec{n}_{j}'}\}=\delta_{\vec{n}_j\vec{n}_{j}'}$.

The field equations \eqref{eqm-frw2} imply that the modes are dynamically decoupled (both if one employs the real or the complex eigenfunctions), with equations of motion
\begin{equation}
\label{meom1}
\ddot{q}_{\vec{n}_j}+ [\omega_{n}^{2}+s(t)]q_{\vec{n}_j}=0,
\end{equation}
and similarly for the $x$'s (i.e., the sine modes). Thus, the dynamical equations of the modes only depend on the corresponding LB eigenvalue $-\omega^{2}_{n}$, which is given by $\omega^{2}_{n}=n_1^{2}+n_2^{2}+n_3^{2}$. From now on, we will employ an abstract notation such that the subindex $n$ will be a positive number designating the order of these eigenvalues. That is, $\omega^{2}_{n}<\omega^{2}_{n'}$ if $n<n'$. It is also worth noticing that the degeneracy $g_{n}$ of each eigenspace (i.e., the number of independent modes with eigenvalue equal to $-\omega_n^2$) presents a complicated dependence on the label $n$ because of accidental degeneracy: there can exist many different tuples $\vec{n}_{j}$ that lead to the same eigenvalues.

In order to discuss the Fock quantization of this scalar field, we introduce a particular complex structure which will serve as starting point in our analysis. This complex structure $J_0$ is chosen as the structure which would be naturally associated with the case of a free, massless KG field. Hence, it is determined entirely by the three-torus metric. We can define $J_0$ in terms of the LB operator as
\begin{equation}
\label{ml-cs}
J_0\left(\begin{array}{c}
\varphi\\ P_{\varphi}
\end{array}\right)=
\left(\begin{array}{cc}
0 & -(-h\Delta)^{-1/2}\\
(-h\Delta)^{1/2} & 0
\end{array}\right)
\left(\begin{array}{c}
\varphi\\ P_{\varphi}
\end{array}\right).
\end{equation}
By construction, then, its invariance under the group of isometries is warranted.

Returning to the mode description, it is convenient to define annihilationlike variables
\begin{equation}
a_{\vec{n}_j}=\frac{1}{\sqrt{2\omega_{n}}}(\omega_{n}q_{\vec{n}_j}+i p_{\vec{n}_j}), \qquad  \tilde{a}_{\vec{n}_j}=\frac{1}{\sqrt{2\omega_{n}}}(\omega_{n}x_{\vec{n}_j}+i y_{\vec{n}_j}),
\end{equation}
as well as the corresponding creationlike variables, given by their complex conjugates $a^{\ast}_{\vec{n}_j}$ and $\tilde{a}^{\ast}_{\vec{n}_j}$.
In the following, we ignore the zero mode (i.e., the mode with $\omega_{n}=0$) in our discussion. This mode can be quantized separately, possibly by nonstandard methods. Note that its exclusion does not alter the properties of the system which have to do with the fieldlike behavior, arising from the presence of an infinite number of degrees of freedom. With this caveat, our annihilation and creationlike variables are well defined. The action of $J_{0}$ on the basis in phase space provided by these variables is diagonal, and takes the standard form $J_{0}(a_{\vec{n}_j})=ia_{\vec{n}_j}$, $J_{0}(a^{\ast}_{\vec{n}_j})=-ia^{\ast}_{\vec{n}_j}$, and likewise for $\tilde{a}_{\vec{n}_j}$ and $\tilde{a}^{\ast}_{\vec{n}_j}$. Therefore, the introduced variables are precisely those which are promoted to annihilation and creation operators in the Fock representation determined by $J_{0}$.

Taking into account the dynamical equations \eqref{meom1} of the modes, the evolution of the annihilation and creationlike variables from the fixed time $t_{0}$ to any another time $t\in \mathbb{I}$ is just a linear transformation $\mathcal{U}(t,t_0)$ which is block diagonal, with $ 2 \times 2$ blocks of the form
\begin{equation}
\label{ecav}
\left(\begin{array}{c}
a_{\vec{n}_j}(t)\\ a_{\vec{n}_j}^{\ast}(t)
\end{array}\right)=
\left(\begin{array}{cc}
\alpha_{n}(t,t_{0}) & \beta_{n}(t,t_{0})\\
\beta_{n}^{\ast}(t,t_{0}) & \alpha_{n}^{\ast}(t,t_{0})
\end{array}\right)
\left(\begin{array}{c}
a_{\vec{n}_j}(t_{0})\\
a_{\vec{n}_j}^{\ast}(t_{0})
\end{array}\right),
\end{equation}
and similarly for $(\tilde{a}_{\vec{n}_j},\tilde{a}_{\vec{n}_j}^{\ast})$. Since the equations of motion \eqref{meom1} depend only on the value of $\omega_n$, the functions $\alpha_{n}(t,t_0)$ and $\beta_{n}(t,t_0)$ are independent of the exact form of $\vec{n}_j$ as far as $\omega^{2}_{n}=|\vec{n}_j|^{2}$. Consequently, all the annihilation and creationlike variables associated with the same eigenspace of the LB operator have the same evolution. Besides, since the dynamical evolution is a symplectic transformation, the alpha and beta functions can then be regarded as Bogoliubov coefficients, and we have that $|\alpha_{n}(t,t_{0})|^{2}-|\beta_{n}(t,t_{0})|^{2}=1$, $\forall n$ and for all times $t$ (and initial times $t_{0}$).

\section{Unitary evolution and uniqueness of the representation}
\label{s3}

In this section, we will show that the Fock representation determined by the complex structure $J_{0}$ leads to a unitary quantum evolution, assuming very mild differentiability conditions on the mass function $s(t)$. In addition, we will prove that every invariant Fock representation with a unitary dynamics is unitarily equivalent to the one defined by $J_{0}$. Therefore, the criteria of symmetry invariance and unitary evolution select a unique class of equivalent Fock representations. Since part of the demonstration is similar to that presented in Ref. \cite{CMV8} for the three-sphere, we will summarize the main steps of the proof and focus our attention on the peculiarities of the three-torus case.

\subsection{Unitary dynamics}
\label{s3-1}

As we have commented, a linear canonical transformation $T$ can be implemented as a quantum unitary transformation in the representation determined by a complex structure $J$ if and only if the corresponding beta Bogoliubov coefficients are square summable, i.e.
\begin{equation}
\label{uiec1}
\sum_{j=0}^{3}\sum_{\vec{n}_{j}} 2|\beta_{\vec{n}_j}(t,t_0)|^{2}=\sum_{n}g_{n}|\beta_{n}(t,t_0)|^{2} < \infty, \qquad \forall t \in \mathbb{I}.
\end{equation}
Here, we have taken into account the two modes corresponding to each value of $\vec{n}_j$ (namely, the sine and cosine modes), and zero modes are already excluded. The asymptotic behavior of the alpha and beta coefficients for large eigenvalues $\omega_n$ of the LB operator (or, equivalently, for large $n$) can be obtained exactly as in  Ref. \cite{CMV8}. Explicitly, this asymptotic behavior is
\begin{equation}
\label{oabf1}
\alpha_{n}(t,t_0)=e^{-i\omega_{n}(t-t_0)}+\mathcal{O}\left(\frac{1}{\omega_{n}}\right), \qquad \beta_{n}(t,t_0)=\mathcal{O}\left(\frac{1}{\omega^{2}_{n}}\right),
\end{equation}
where the symbol $\mathcal{O}$ indicates the asymptotic order. In order to deduce this behavior, it suffices to impose that the mass function $s(t)$ have a first derivative which is integrable in every closed subinterval of $\mathbb{I}$. Using this result, it is straightforward to see that condition \eqref{uiec1} is satisfied when the sequence $\{g_{n}/\omega^{4}_{n}\}$ is square summable (over $n$). To check this summability, it is necessary to study the behavior of the degeneracy $g_{n}$ for large $n$. In fact, the variation of $g_n$ with $n$ is much more complicated than, e.g., in the case of the three-sphere, studied in Ref. \cite{CMV8}. Owing to accidental degeneracy, the exact dependence is not explicitly available. Nonetheless, it is still possible to compute the asymptotic behavior in which we are interested. From the expression of the eigenvalue $\omega_{n}$, it is clear that the set of positive integers $\{m;\, m\in \mathbb{N}^+\}$ is a subset of the possible values of $\omega_{n}$. Taking 
into account that $1/\omega^{4}_{n}$ is a decreasing function of $\omega_{n}$ and that $g_n$ is strictly positive, one can then easily realize that the summability condition on the sequence $\{g_{n}/\omega^{4}_{n}\}$ is satisfied if the sequence $\{D_{m}/m^{4}\}$ is summable (over $m$), where $D_{m}$ is the number of eigenstates whose eigenvalue $\omega_{n}$ belongs to the interval $(m,m+1]$. The computation of $D_{m}$ can be done as in the analysis of the classical black body spectral radiation that leads to the Rayleigh-Jeans law (see e.g. Ref. \cite{blackbody}). Note that every tuple $(n_1,n_2,n_3)$, where $n_{i}\in \mathbb{Z}$, determines uniquely an LB eigenstate with eigenvalue $\omega^{2}_{n}=n_{1}^{2}+n_{2}^{2}+n_{3}^{2}$. The two tuples with opposite signs for all $n_i$ can be assigned to the sine and cosine modes with the same label $\vec{n}_j$. These tuples form a cubical lattice of step equal to one in $\mathbb{R}^{3}$. The corresponding ``average eigenstate density'', defined in $\mathbb{R}^{3}$,
 is just the unity. The number of eigenstates contained in a sphere of radius equal to $m$, which we will call $N_{m}$, approaches the sphere volume for large $m$, namely, $N_{m} \simeq 4\pi m^{3}/3$. On the other hand, we clearly have $D_{m}=N_{m+1}-N_{m}$. Then, $D_{m} \propto m^{2}$ when $m \rightarrow \infty$. Consequently, the sequence $\{D_{m}/m^{4}\}$ is summable, and condition \eqref{uiec1} is satisfied. Therefore, the evolution of the system becomes a quantum unitary transformation in the Fock representation specified by $J_0$.

\subsection{Uniqueness of the Fock representation: Invariant representations}

We will show now that all invariant Fock representations with unitary dynamics are equivalent to the representation obtained from $J_{0}$. We start by characterizing the complex structures that are invariant under the group of symmetries of the three-torus, formed by the transformations $T_{\vec{\alpha}}$. To reach this characterization, we will adopt an approach that differs from that followed in the case of three-sphere topology, studied in detail in Ref. \cite{CMV8}. While, for the three-sphere, each of the eigenspaces of the LB operator provides a distinct irreducible representation of the rotation group $SO(4)$, the symmetries of the three-torus form an Abelian group, and therefore its irreducible representations are all one-dimensional and defined on \emph{complex} vector spaces. The eigenspaces of the LB operator now aquire accidental degeneracy, each of them splitting into a finite number of irreducible representations of the symmetry group. Besides, as already mentioned, the structure of the 
degeneracy is rather complicated, and a closed expression for the dimension of the eigenspaces is not even available. These obstacles are not encountered in the case of the three-sphere studied in Refs. \cite{CMV8,CMOV-S3S,CMOV-FTC}, where the analysis is considerably simplified by the application of the Schur's lemma. Hence, the case of the three-torus calls for a detailed analysis. We will present two different but completely equivalent ways to deal with the peculiarities that arise in this characterization: one of them uses a basis of real modes to expand the field, whereas the other one goes along with complex modes. In this subsection, we determine the characterization in terms of real modes. The analysis employing complex modes and the transformation between the complex and real bases can be found in the Appendix.

Let us study the real decomposition defined in Eq. \eqref{rfd1}. A complex structure $J$ is invariant under a transformation $T_{\vec{\alpha}}$ if and only if $T_{\vec{\alpha}}^{-1}JT_{\vec{\alpha}}=J$. On the other hand, the action of $T_{\vec{\alpha}}$ on the field modes can be easily derived from the (active) transformation of the field using Eq. \eqref{rfd1}. This yields
\begin{equation}
\label{rmt1}
q'_{\vec{n}_j}=\cos(\vec{n}_j\cdot\vec{\alpha})q_{\vec{n}_j}-\sin(\vec{n}_j\cdot\vec{\alpha})x_{\vec{n}_j}, \qquad x'_{\vec{n}_j}=\sin(\vec{n}_j\cdot\vec{\alpha})q_{\vec{n}_j}+\cos(\vec{n}_j\cdot\vec{\alpha})x_{\vec{n}_j}.
\end{equation}
One obtains similar equations for the momentum field modes $(p_{\vec{n}_j},y_{\vec{n}_j})$.

Notice that these transformations do not mix all modes, not even all those in the same eigenspace of the LB operator --as occurred e.g. in the three-sphere case \cite{CMV8}. The mixing happens only between the pair of modes with the same label $\vec{n}_j$. The corresponding invariant subspaces under the symmetry group are therefore two dimensional, which is in fact the minimum possible dimensionality of any real unitary representation of a (nontrivial) Abelian compact group \cite{brocker}. In this sense, the isometry group of the three-torus corresponds to the most extreme case.

Actually, the situation is equivalent to that found in the case of the circle, $S^{1}$ (studied in the dimensional reduction of the linearly polarized Gowdy cosmologies). Thus, the characterization of the invariant complex structures can be made following a line of arguments similar to that presented in Ref. \cite{ccmv1}. It is not difficult to realize from Eq. \eqref{rmt1} that $T_{\vec{\alpha}}$ acts both on the variables $\{(q_{\vec{n}_j},x_{\vec{n}_j})\}$ and on $\{(p_{\vec{n}_j},y_{\vec{n}_j})\}$ as the product of three copies of $SO(2)$, one for each direction. In fact, one obtains a different (real) representation of this group for each value of the $\vec{n}_j$ label. As a consequence, all invariant complex structures must be block diagonal, with $4\times4$ blocks, denoted by $J_{\vec{n}_j}$. In addition, each block $J_{\vec{n}_j}$ is composed of four $2\times2$ blocks: $J_{\vec{n}_j}^{qq}$, $J_{\vec{n}_j}^{pq}$, $J_{\vec{n}_j}^{qp}$, and $J_{\vec{n}_j}^{pp}$, which, owing to the 
invariance
condition $T_{\vec{\alpha}}^{-1}JT_{\vec{\alpha}}=J$, must be matrices with a real skew-symmetric part and with a symmetric part proportional to the identity. These blocks define maps between the configuration ($q$) and momentum ($p$) sectors of phase space, as indicated by the introduced superindex notation.

We must still demand the remaining conditions that the square of the complex structure be minus the identity, that it must preserve the symplectic structure, and that it must be compatible with it (namely, it must provide a definite positive bilinear map when composed with it). Let us first impose the condition that the bilinear map $\Omega(J\cdot, \cdot)$ be positive definite on phase space. This means that $J_{\vec{n}_j}^{T}\Omega_{\vec{n}_j}$ must be a symmetric positive definite matrix, for all values of the labels $\vec{n}_j$. Here, the superindex $T$ denotes the transpose matrix, and the blocks $\Omega_{\vec{n}_j}$
of the symplectic structure [in the real mode basis
$\{(q_{\vec{n}_j},x_{\vec{n}_j},p_{\vec{n}_j},y_{\vec{n}_j})\}$] are
\begin{equation}
\label{njss}
\Omega_{\vec{n}_j}=\left(\begin{array}{cc} \mathbf{0}_{2\times2} & -\mathbf{I}_{2\times2}\\ \mathbf{I}_{2\times2} & \,\,\,\mathbf{0}_{2\times2}
\end{array}
\right).
\end{equation}
The matrices $\mathbf{0}_{2\times2}$ and $\mathbf{I}_{2\times2}$ are the zero and the identity $2\times 2$ matrices. One can see that the imposed positivity condition implies that $J_{\vec{n}_j}^{pq}$ and $-J_{\vec{n}_j}^{qp}$ must be symmetric positive definite matrices, and that $J_{\vec{n}_j}^{qq}=-J_{\vec{n}_j}^{pp}$. Together with our previous results, we get that $J_{\vec{n}_j}^{pq}=c_{\vec{n}_j}^{(1)}\,\mathbf{I}_{2\times2}$ and $J_{\vec{n}_j}^{qp}=-c_{\vec{n}_j}^{(2)}\,\mathbf{I}_{2\times2}$, with $c_{\vec{n}_j}^{(1)},c_{\vec{n}_j}^{(2)}>0$.

Finally, the remaining conditions on the complex structure translate into the following requirements on matrices:
$J_{\vec{n}_j}^{T}\Omega_{\vec{n}_j}J_{\vec{n}_j}=\Omega_{\vec{n}_j}$ (i.e., the complex structure is a symplectic transformation) and
$J_{\vec{n}_j}^{2}=-\mathbf{I}_{4\times4}$ ($J^2$ is minus the identity). From these equations, one concludes that
$J_{\vec{n}_j}^{qq}=-J_{\vec{n}_j}^{pp}=\pm
c_{\vec{n}_j}^{(3)}\,\mathbf{I}_{2\times2}$ with  $c_{\vec{n}_j}^{(3)}=\sqrt{c_{\vec{n}_j}^{(1)}c_{\vec{n}_j}^{(2)}-1}$, so that, in particular, we must have $c_{\vec{n}_j}^{(1)}c_{\vec{n}_j}^{(2)}\ge1$.

Summarizing, we get that every invariant complex structure
must be block diagonal in our basis of real modes, with $4\times4$ blocks of
the form
\begin{equation}
J_{\vec{n}_j}=\left(\begin{array}{cc}
\pm c_{\vec{n}_j}^{(3)}\,\mathbf{I}_{2\times2} & -c_{\vec{n}_j}^{(2)}\,\mathbf{I}_{2\times2} \\ c_{\vec{n}_j}^{(1)}\,\mathbf{I}_{2\times2} & \mp c_{\vec{n}_j}^{(3)}\,\mathbf{I}_{2\times2}
\end{array}\right),
\end{equation}
where $c_{\vec{n}_j}^{(1)}$ and $c_{\vec{n}_j}^{(2)}$ are positive numbers such that $c_{\vec{n}_j}^{(1)}c_{\vec{n}_j}^{(2)}\ge1$, and $c_{\vec{n}_j}^{(3)}=\sqrt{c_{\vec{n}_j}^{(1)}c_{\vec{n}_j}^{(2)}-1}$.
It is worth noting that the deduced proportionality to the identity of the above $2\times2$ blocks arises as a consequence of all the conditions imposed to guarantee that $J$ is a compatible invariant complex structure (in this aspect, our discussion does not parallel those of Refs. \cite{CMV8,CMOV-S3S,CMOV-FTC}).

We now introduce a change of basis, passing to the basis formed by the annihilation and creationlike variables
$\{(a_{\vec{n}_j},a_{\vec{n}_j}^{\ast},\tilde{a}_{\vec{n}_j},\tilde{a}_{\vec{n}_j}^{\ast})\}$, where the expression of $J_{0}$ is especially simple.
Every compatible invariant complex structure $J$ results then block diagonal, with $2\times 2$ blocks that coincide by pairs, namely
\begin{equation}
J_{\vec{n}_j}=\left(\begin{array}{cc}
\mathcal{J}_{\vec{n}_j} & \mathbf{0}_{2\times2}\\ \mathbf{0}_{2\times2} &  \mathcal{J}_{\vec{n}_j}
\end{array} \right), \end{equation}
with
\begin{equation}
\mathcal{J}_{\vec{n}_j}= i\left(\begin{array}{cc}
c_{\vec{n}_j}^{(2)} + c_{\vec{n}_j}^{(1)}\omega^{2}_{n} &
c_{\vec{n}_j}^{(2)}-c_{\vec{n}_j}^{(1)}\omega^{2}_{n}\pm i c_{\vec{n}_j}^{(3)}\omega_{n}\\
-c_{\vec{n}_j}^{(2)}+c_{\vec{n}_j}^{(1)}\omega^{2}_{n}\pm i
c_{\vec{n}_j}^{(3)}\omega_{n} &
-c_{\vec{n}_j}^{(2)} - c_{\vec{n}_j}^{(1)}\omega^{2}_{n}
\end{array}
\right).
\end{equation}
Actually, every invariant complex structure $J$ can be obtained from $J_0$ by means of a symplectic transformation, $K$, namely $J=KJ_0K^{-1}$ \cite{ccmv1}. Since $J_0$ adopts the diagonal form $(J_0)_{\vec{n}_j}=\text{diag}\{i,-i,i,-i\}$, the symplectic transformation $K$ can also be taken block diagonal, with $4\times4$ blocks of the following type:
\begin{equation}
\label{st1}
K_{\vec{n}_j}=\left(\begin{array}{cc}
\mathcal{K}_{\vec{n}_j} & \mathbf{0}_{2\times2} \\ \mathbf{0}_{2\times2} & \mathcal{K}_{\vec{n}_j}
\end{array}\right), \qquad
\mathcal{K}_{\vec{n}_j} = \left(\begin{array}{cc}
\kappa_{\vec{n}_j} & \lambda_{\vec{n}_j} \\
\lambda_{\vec{n}_j}^{\ast} & \kappa_{\vec{n}_j}^{\ast}
\end{array}\right).
\end{equation}
Here, $\kappa_{\vec{n}_j}$ and $\lambda_{\vec{n}_j}$ are complex numbers playing the role of alpha and beta Bogoliubov coefficients for $K$, so that $|\kappa_{\vec{n}_j}|^{2}-|\lambda_{\vec{n}_j}|^{2}=1$ $\forall \vec{n}_j$. They are related with the constants $c_{\vec{n}_j}^{(i)}$ by the equations
\begin{equation}
|\kappa_{\vec{n}_j}|^2+|\lambda_{\vec{n}_j}|^{2}=c_{\vec{n}_j}^{(2)} + c_{\vec{n}_j}^{(1)}\omega^{2}_{n}, \qquad
-2  \kappa_{\vec{n}_j}\lambda_{\vec{n}_j}=c_{\vec{n}_j}^{(2)}-c_{\vec{n}_j}^{(1)}\omega^{2}_{n}\pm i c_{\vec{n}_j}^{(3)}\omega_{n}.
\end{equation}

\subsection{Uniqueness of the Fock representation: Unitary dynamics}
\label{s3-3}

We are interested only in those invariant complex structures $J$ that allow for a unitary implementation of the dynamics of the system. Since $J$ and $J_0$ are related by a symplectomorphism $K$ (which can be regarded as a change of the basis of annihilation and creationlike variables), we get that a symplectic transformation $T$ can be implemented unitarily with respect to the former of these complex structures if and only if the ``mirror'' transformation $K^{-1}TK$ is unitarily implementable with respect to $J_0$ \cite{ccmv1}. Specializing the discussion to the case of the dynamics \eqref{ecav}, we obtain a ``new'' map $K^{-1}UK$ which can be seen as the evolution of the annihilation and creationlike variables corresponding to the complex structure $J$ (but expressed in the basis associated with $J_0$). The new beta coefficients, that we will call $\beta_{\vec{n}_j}^{J}$, can be written in terms of the original Bogoliubov coefficients as
\begin{equation}
\label{nbac1}
\beta_{\vec{n}_j}^{J}(t,t_0)=(\kappa_{\vec{n}_j}^{\ast})^{2}\beta_{n}(t,t_0)-
(\lambda_{\vec{n}_j})^{2}\beta_{n}^{\ast}(t,t_0)+2i\kappa_{\vec{n}_j}^{\ast}\lambda_{\vec{n}_j}\mathfrak{I}[\alpha_{n}(t,t_0)],
\end{equation}
where $\mathfrak{I}[\cdot]$ denotes the imaginary part.

The unitary implementability condition is again the square summability --as in Eq. \eqref{uiec1}-- of the new set of beta coefficients. Thus, we arrive at a situation very similar to that studied in Ref. \cite{CMV8}. Following the arguments explained in that reference, recalling the proven unitarity of the dynamics with respect to $J_0$, and taking into account that $|\kappa_{\vec{n}_j}|\ge1$  $\forall \vec{n}_j$, it is not difficult to realize that, if the set given by $\{\beta_{\vec{n}_j}^{J}(t,t_0)\}$ is indeed square summable, then the same must be true for $\{z_{\vec{n}_j}\mathfrak{I}[\alpha_{n}(t,t_{0})]\}$, with $z_{\vec{n}_j}=\lambda_{\vec{n}_j}/\kappa_{\vec{n}_j}^{\ast}$. Given the asymptotic behavior of $\alpha_{n}(t,t_0)$ for large $n$, which is also studied in Ref. \cite{CMV8}, this condition turns out to imply the square summability of the set
\begin{equation}
\label{ssc1}
\left\{z_{\vec{n}_j}\sin\left[\omega_{n}(t-t_0)+\int_{t_{0}}^{t}d\bar{t}\,\frac{s(\bar{t})}{2\omega_{n}}\right] \right\}.
\end{equation}
In order to reach this last result, we have supposed (as a sufficient condition) that the mass function $s(t)$ has a second derivative which is integrable in every compact subinterval of the time domain $\mathbb{I}$. Then an integration in time, over any such subinterval, of the (partial) square sum(s) of the set \eqref{ssc1} and an appropriate application of Luzin's theorem \cite{Luzin} (which is possible because the elements of the analyzed set are measurable functions) show that the considered property of square summability necessarily implies the same for the set $\{z_{\vec{n}_j}\}$. From this, it is not difficult to deduce as well the square summability of the set $\{\lambda_{\vec{n}_j}\}$, using the definition of $z_{\vec{n}_j}$ and that $|\kappa_{\vec{n}_j}|^{2}=1+|\lambda_{\vec{n}_j}|^2$\cite{CMV8}. But this last summability is precisely the condition for the equivalence of the Fock representations determined by $J$ and $J_0$, since it amounts to the unitarity of the symplectic transformation that
relates the two complex structures. In summary, every Fock representation which is invariant under the three-torus symmetries and allows for a unitary evolution turns out to be unitarily equivalent to the Fock representation determined by $J_0$.

\section{Uniqueness of the field description}
\label{s4}

Once proven that our criteria remove the ambiguity in the choice of representation of the CCR's for a field satisfying Eq. \eqref{fe1}, we will discuss in this section the additional ambiguity associated with the selection of a canonical pair of field variables, among all those related by a time dependent linear canonical transformation in which the field configuration is scaled. As we have mentioned, this ambiguity arises naturally in cosmological contexts, in which it is common to scale the field by a time dependent function, absorbing in this way part of the variation due to the (background, effective, or auxiliary) spacetime in which the field propagates. Although this kind of canonical changes lead to equivalent classical descriptions, this is not true in general from a quantum viewpoint, both because not all linear canonical transformations can be implemented unitarily in the quantum theory and because the field dynamics is altered by the time dependence of the transformation, thus affecting the
possible unitarity of the evolution.

Let us consider the family of time dependent linear canonical transformations obtained by completing a time dependent scaling of the field into a symplectic map:
\begin{equation}
\label{tds1}
\phi=f(t)\varphi, \qquad P_{\phi}=\frac{P_{\varphi}}{f(t)}+g(t)\sqrt{h}\varphi.
\end{equation}
Note that the inverse scaling of the momentum, needed for the transformation to be canonical, has been complemented with a (conveniently densitized) time dependent linear contribution of the field configuration. This contribution is clearly allowed in the most general linear canonical transformation, and it is necessary if one wants to avoid (problematic) cross terms between the field configuration and the field momentum in the Hamiltonian of the transformed system. Here, the real functions $f(t)$ and $g(t)$ are supposed to be, at least, twice differentiable in order to preserve the differential formulation of the field theory. In addition, with the aim at preventing the introduction of (new) singularities, $f(t)$ must be a nonvanishing function. As explained in Ref. \cite{cmv}, we can assume, without loss of generality, that $f(t_0)=1$ and $g(t_0)=0$, because every other initial set of data can be obtained with an additional canonical transformation like \eqref{tds1}, but with constant coefficients instead
of time dependent functions. Actually, the Fock representations of the new canonical pair of field variables reached with this constant transformation can be constructed from those of the original pair simply by linearity, and therefore are the same. This still applies when one imposes our criteria of symmetry invariance and unitary dynamics, which are stable under constant linear canonical transformations \cite{cmv}.

We start by introducing the same type of Fock representation for the new canonical pair $(\phi,P_{\phi})$ as we adopted for the original pair $(\varphi, P_{\varphi})$, i.e. the representation determined by the complex structure $J_0$. The new Bogoliubov coefficients  $\big(\tilde{\alpha}_{n}(t,t_0), \tilde{\beta}_{n}(t,t_0)\big)$ that describe the evolution of the annihilation and creationlike variables obtained from the new canonical pair can be written in terms of the original coefficients, given in Eq. \eqref{oabf1}, as (see Ref. \cite{CMOV-S3S})
\begin{eqnarray}
\label{nabf1}
\tilde{\alpha}_{n}(t,t_0)=f_{+}(t)\alpha_{n}(t,t_0)+f_{-}(t)\beta_{n}^{\ast}(t,t_0)
+i\frac{g(t)}{2\omega_{n}}[\alpha_{n}(t,t_{0})+\beta_{n}^{\ast}(t,t_0)],\\ \label{nabf2}
\tilde{\beta}_{n}(t,t_0)=f_{+}(t)\beta_{n}(t,t_0)+f_{-}(t)\alpha_{n}^{\ast}(t,t_0)
+i\frac{g(t)}{2\omega_{n}}[\beta_{n}(t,t_{0})+\alpha_{n}^{\ast}(t,t_0)],
\end{eqnarray}
where we have defined $2f_{\pm}(t)=f(t)\pm1/f(t)$. Just as in the previous section for the original pair of fields, every invariant compatible complex structure $J$, providing an invariant Fock representation for the new canonical pair, can be obtained from $J_0$ via a symplectic transformation $K$ with the form \eqref{st1}. Therefore, with respect to any such a complex structure $J$, the beta coefficients associated with the pair $(\phi,P_{\phi})$ take the expression
\begin{equation}
\label{nbJc1}
\tilde{\beta}_{\vec{n}_j}^{J}(t,t_0)=(\kappa_{\vec{n}_j}^{\ast})^{2}\tilde{\beta}_{n}(t,t_0)
-(\lambda_{\vec{n}_j})^{2}\tilde{\beta}_{n}^{\ast}(t,t_0)
+2i\kappa_{\vec{n}_j}^{\ast}\lambda_{\vec{n}_j}\mathfrak{I}[\tilde{\alpha}_{n}(t,t_0)].
\end{equation}
Again, the set given by $\{\tilde{\beta}_{\vec{n}_j}^{J}(t,t_0)\}$ must be square summable at all times in the interval $\mathbb{I}$ to obtain a unitary evolution.

To make more transparent the rest of our discussion, it is convenient to change our notation now and relabel the beta (and alpha) coefficients --as well as the kappa and lambda coefficients-- as $\{\tilde{\beta}^{J}_{n,l}(t,t_{0})\}$, where we recall that $n$ is the index characterizing the eigenvalue of the LB operator, while $l=1,..., g_n$ takes care of the corresponding eigenspace degeneracy, i.e., the different values of $\vec{n}_j$ contained in that eigenspace, each with an additional double degeneracy owing to the presence of sine and cosine modes. We will prove that a unitary quantum evolution is possible only if $f(t)=1$ and $g(t)=0$ at all times, namely, only for the original pair of canonical field variables. In order to demonstrate that $f(t)=1$, we will follow an adaptation of the procedure explained in Ref. \cite{CMOV-S3S} which incorporates the peculiarities of the spectrum of the LB operator in the case of the three-torus. On the other hand, the proof that $g(t)=0$ requires a line of 
arguments which differs from that presented in Ref. \cite{CMOV-S3S} (see Ref. \cite{CMOV-GUS}).

Let us consider the sequences $\{\tilde{\beta}^{J}_{n, M_n}(t,t_0)\}$, obtained from the set $\{\tilde{\beta}^{J}_{n,l}(t,t_0)\}$ by choosing for each $n$ just one (arbitrary) element $M_n\in\{l\}$. It is clear that the square summability condition on $\{\tilde{\beta}^{J}_{n,l}(t,t_0)\}$ leads to the square summability of the sequence $\{\tilde{\beta}^{J}_{n, M_n}(t,t_0)\}$ at all times. Substituting the asymptotic behavior \eqref{oabf1} for $\alpha_{n}(t,t_0)$ and $\beta_{n}(t,t_0)$ in expression \eqref{nbJc1} [taking into account Eqs. \eqref{nabf1} and \eqref{nabf2}], and employing that $|\kappa_{n,M_{n}}|\ge 1$ ($\forall n,M_{n}$), it is straightforward to see that a necessary condition for the square summability of the studied sequence is that the terms
\begin{equation}
\label{ee1}
\left[e^{i\omega_n(t-t_0)}-z^2_{n,M_{n}}e^{-i\omega_n(t-t_0)}\right]f_-(t)-2iz_{n,M_{n}}\sin[\omega_{n}(t-t_0)]f_{+}(t)
\end{equation}
tend to zero at large $n$, $\forall t \in \mathbb{I}$.

On the other hand, let us note that, given any particular eigenvalue $\omega_{(0)}$ of the LB operator, every real number $m\omega_{(0)}$, with $m\in\mathbb{N}^+$, is also an eigenvalue, obtained by a scaling in each direction with the same factor $m$. Since the terms \eqref{ee1} must tend to zero when $\omega_n\rightarrow\infty$ ($n\rightarrow\infty$), the subsequence constructed by restricting $\omega_{n}$ to the subset $m\omega_{(0)}$ must also have a vanishing limit when $m \rightarrow \infty$. Now, considering a specific subset of times $t=t_0+2\pi r/\omega_{(0)}$, where $r$ can be any positive integer satisfying that $t\in \mathbb{I}$, and calling $z_{m\omega_{(0)}}$ the value of $z_{n,M_{n}}$ when $\omega_{n}=m\omega_{(0)}$, we obtain from the real and imaginary parts of Eq. \eqref{ee1} that both
\begin{equation}
\left(1-\mathfrak{R}\left[z_{m\omega_{(0)}}^{2}\right]\right)f_{-}\!\left(t_{0}+\frac{2\pi r}{\omega_{(0)}}\right), \qquad \mathfrak{I}\left[z_{m\omega_{(0)}}^{2}\right]f_{-}\!\left(t_{0}+\frac{2\pi r}{\omega_{(0)}}\right),
\end{equation}
have to vanish when $m$ goes to infinity, for every possible value of $r$ and $\omega_{(0)}$. Here, $\mathfrak{R}[\cdot]$ stands for the real part. In Refs. \cite{cmv,CMOV-S3S}, it is proven that, if the terms \eqref{ee1} tend to zero, then $1-\mathfrak{R}[z_{m\omega_{(0)}}^{2}]$ and $\mathfrak{I}[z_{m\omega_{(0)}}^{2}]$ cannot have simultaneously a zero limit at large $m$ in any (sub)sequence, and therefore the only possibility left is that $f_{-}(t_{0}+2\pi r/\omega_{(0)})$ vanishes, $\forall r$ and $\forall \omega_{(0)}$. Taking into account the continuity of the function $f(t)$, that $f(0)=1$, and that the subset $\{t_{0}+2\pi r/\omega_{(0)}\}$ is dense in $\mathbb{I}$, we then conclude that $f(t)$ is necessarily the unit function in the whole considered time interval. This fixes any field scaling to the trivial one.

Let us finally prove that the function $g(t)$, which controls the possible redefinition of the momentum by a linear contribution of the field configuration, must also vanish if the dynamics is unitary. We consider again the square summability condition on the set of beta coefficients $\{\tilde{\beta}^{J}_{n,l}(t,t_0)\}$ at all instants of time, which amounts to the unitary implementation of the evolution, now specialized to the case in which $f(t)$ is the unit function. Obviously, a direct consequence is the square
summability of the set $\{\tilde{\beta}^{J}_{n,l}(t,t_0)/(\kappa_{n,l}^{\ast})^{2}\}$, since $\kappa_{n,l}$ is bounded in norm from below by the unit. Recalling that the sequence $\{\sqrt{g_{n}}\beta_{n}(t,t_{0})\}$ is square summable (because the dynamics is unitary with respect to $J_0$), the asymptotic behavior of $\alpha_{n}(t,t_0)$ given in Eq. \eqref{oabf1}, and that $|z_{n,l}|\le1$, it is not difficult to check then the square summability of the set $\{A_{n,l}(t,t_0)\}$ at all instants of time, where
\begin{equation}
A_{n,l}(t,t_0)=2|z_{n,l}|\mathfrak{I}\left[\alpha_{n}(t,t_0)\right]+\frac{g(t)}{
2\omega_{n}} \left[e^{i(\omega_{n}\tau-\delta_{n,l})}+|z_{n,l}|^{2}e^{-i(\omega_{n}\tau-\delta_{n,l})}+2|z_{n,l}|
\cos(\omega_{n}\tau)\right].\end{equation}
Here, we have written $z_{n,l}=|z_{n,l}|e^{i\delta_{n,l}}$, and $\tau=t-t_0$. Let us consider now the set
$\{A_{n,l}(t,t_0)/\omega_{n}\}$, that must be also square summable since $\omega_n\rightarrow \infty$.
Hence, employing the summability of the sequence $\{g_{n}/\omega^{4}_{n}\}$, proven in  Sec. \ref{s3-1}, it follows that the
set $\{|z_{n,l}|\mathfrak{I}\left[\alpha_{n}(t,t_0)\right]/\omega_{n}\}$ must be square summable. This implies that
$\{|z_{n,l}|/\omega_{n}\}$ must be square summable as well (as can be demonstrated by means of an integration over the time domain and a convenient use of Luzin's theorem, similar to that commented in Sec. \ref{s3-3}).

Taking into account this result and the expression of $A_{n,l}(t,t_0)$, we arrive at the
square summability (at all times) of the set formed by
\begin{equation}
B_{n,l}(t,t_0)=2|z_{n,l}|\mathfrak{I}\left[\alpha_{n}(t,t_0)\right]+\frac{g(t)}{
2\omega_{n}}e^{ i(\omega_{n}\tau-\delta_{n,l})}.
\end{equation}
By considering the imaginary part of these quantities, which form a square summable set a fortiori, we conclude that either $g(t)$ vanishes identically  or the sequence $\{g_{n}/\omega^{2}_n\}$ must be summable (it suffices to perform again a suitable integration in time in any subinterval where the continuous function $g(t)$ would differ from zero \cite{CMOV-GUS}).
However, this last possibility is ruled out, because the sum of $g_{n}/\omega^{2}_n$ diverges [this sum is greater than the sum of $D_{n}/(n+1)^2$ over the positive integers, and the later clearly diverges, given the asymptotic behavior $D_{n} \propto n^{2}$ discussed in Sec. \ref{s3-1}].
In conclusion, $g(t)$ has to be the zero function.

In total, we have shown that a Fock quantization with a vacuum invariant under the three-torus isometries and with unitary evolution can be obtained only for the canonical pair satisfying the field equations \eqref{fe1}: no time dependent scaling or redefinition of the field momentum is allowed.

\section{Discussion and conclusions}
\label{s5}

We have discussed two types of ambiguities that affect the Fock quantization of scalar fields in cosmological scenarios, leading to unitarily inequivalent quantum theories and preventing the extraction of robust predictions. One of these ambiguities is related with the choice of a canonical pair for the field. When one considers a scalar field propagating in a nonstationary spacetime (either a genuine background, an effective spacetime, or an auxiliary spacetime), it is common to introduce a suitable scaling of the field in order to extract part of its time dependence, absorbable in terms of the spacetime variation. Generally, this scaling allows one to reach a field description with a simpler and better behavior \cite{mukhanov}. The scaling can always be reformulated as a time dependent linear canonical transformation, permitting also the inclusion of a time varying, linear contribution of the field configuration to the momentum. This contribution is necessary if one wants the new momentum to be
proportional to the time derivative of the field configuration, at the level of the classical Hamiltonian equations. Although, from a classical viewpoint, the field descriptions corresponding to the original and the new canonical pairs are equivalent, this is not generically the case under quantization. Therefore, it is important to find criteria that remove this ambiguity and determine a sole choice of configuration and momentum variables. The other type of ambiguity that we have analyzed concerns the selection of a specific representation for the CCR's of the chosen canonical pair, among the infinitely many inequivalent possibilities. This ambiguity is well known in quantum field theory \cite{wald}. Essentially, the freedom in the representation is captured in the structure that permits the complexification of the space of solutions and the construction of the one-particle Hilbert space from it, namely, the complex structure. The choice of a complex structure amounts to the choice of a vacuum for the Fock 
representation. Hence, the criteria that we are seeking must also provide a unique vacuum for the particular field description picked out for our system.

More specifically, we have studied the Fock quantization of a scalar field which, after a convenient time dependent scaling, satisfies a KG equation with a time varying mass in a static (background, auxiliary, or effective) spacetime whose spatial sections have three-torus topology. We have shown that both of the considered types of ambiguities are removed by imposing two requirements on the Fock quantization. As a first requirement, we impose the invariance of the vacuum under the symmetries of the three-torus (and therefore the invariance of the complex structure). The second requirement is the unitary implementation of the field dynamics. These requirements respond to quite reasonable physical expectations on the properties of the quantum theory. In particular, renouncing to a unitary evolution leads to problems with the most conventional probabilistic interpretation of quantum mechanics. It is worth noting that, in order to obtain these uniqueness results on the Fock quantization, only a mild assumption 
has been made on the time varying mass function, namely, that it must have a second derivative integrable in every compact time subinterval. Actually, this assumption is only a sufficient condition, not even a necessary one. This conclusion applies to every possible time domain for the field evolution, as far as it is an interval of the real line; in particular, it can be bounded.

Our results can be viewed as a specialization of the discussion for the case of general compact topology carried out in Ref. \cite{CMOV-FTC} inasmuch as the choice of a Fock representation is concerned, once the CCR's are determined. However, in the completely general scenario considered in Ref. \cite{CMOV-FTC}, the physical meaning of the spatial symmetries may not be readily available, then leaving open the question of whether the transformations relating modes in the same eigenspace of the LB operator have a physically relevant interpretation or not. Moreover, the proof presented in Ref. \cite{CMOV-FTC} uses that, when expressing each of these eigenspaces in terms of irreducible representations of the symmetry group, such representations differ in distinct eigenspaces (either by assumption or by construction of the considered symmetry transformations). This should be so if the invariant complex structures are to respect the dynamical decoupling of modes. In the case of the three-torus here discussed, we
have given explicitly the group of spatial isometries, motivated its physical relevance, and shown the role played by them in the restriction of the possible complex structures. Furthermore, we have seen in detail how the eigenspaces of the LB operator are decomposed into direct sums of irreducible representations of the symmetry group, which are all different. Another important point that we have been able to discuss is the interplay between the complex irreducible representations of this group (which are precisely those which are considered in the application of the Schur's lemma when characterizing the invariant complex structures \cite{CMOV-FTC}) and the real ones, which are those that arise naturally in the study of real fields. Finally, other aspects that we have been able to analyze with due care (while in the case of general topology the discussion remains more abstract) are the growth of the number of eigenmodes of the LB operator when the eigenvalue increases (in norm) \cite{CMOV-FTC}, and the fact
that the eigenvalues for the three-torus admit a scaling by any
integer number, a property which allows us to simplify the proof that no time dependent scaling of the field configuration is permitted \cite{CMOV-GUS}.

The importance of our results lies on their application to physically relevant cosmological systems, capable to describe the observed Universe, at least in a first perturbative approximation which includes cosmological (scalar) inhomogeneities \cite{mukh}. The current observations of the Universe suggest that its large scale structure is (approximately) isotropic and homogeneous. Moreover, these observations favor cosmological spacetimes with flat spatial sections \cite{wmap}. Therefore, at large scale, the Universe can be described by a flat FRW spacetime. If one considers a sufficiently large compactification scale, or disregards scales beyond a cosmological one related with the Hubble radius, assuming that they have no physically relevant influence, it is possible to adopt compact spatial sections isomorphic to a three-torus. Therefore, our results can be applied to the quantization of scalar fields and scalar perturbations on realistic flat FRW spacetimes, where the KG field dynamics studied in this work
can be obtained by means of a scaling of the fields by time dependent background functions. The simpler system where this can be achieved is the quantization of a massive, minimally coupled scalar field propagating in a flat FRW spacetime. As it is briefly shown in
Sec. \ref{s1}, a convenient scaling leads to a field equation of the KG type in conformal time.

A more interesting application is found in the quantization of scalar cosmological perturbations in a flat (and compact) FRW model with a massive scalar field . This inflationary system can be studied in a similar way as the system considered in Ref. \cite{fmov} (in that reference, the spatial sections have the topology of a three-sphere). After a suitable gauge fixing, it is possible to successfully apply our uniqueness results, in the scaling proposed by Mukhanov for the scalar perturbations \cite{mukhanov}. In addition, for this system, it is possible to attain a canonical pair of gauge invariant Bardeen potentials \cite{bar} whose evolution is unitarily implementable in the privileged Fock representation selected by our criteria \cite{fmov}. These gauge invariants are the energy density and matter velocity perturbations \cite{bar}. Actually, in conformal time and with a suitable scaling, the gauge invariant energy density perturbation amplitude is known to satisfy a second order linear differential
equation of the KG type, with a time dependent mass term \cite{mukhanov,bar}. If one starts from such a KG equation, the unique Fock quantization that one selects with our criteria is in fact unitarily equivalent to the one obtained from the gauge fixed system, as required for consistency \cite{fmov}. Remarkably, in this gauge fixed system, the Bardeen potentials are defined in terms of the original perturbations by means of a time dependent canonical transformation which is nonlocal (because the canonical transformation is different for different modes). In this sense, it is reassuring that one arrives at the commented result of unitary equivalence. The proof that this equivalence is in fact a property not just of the two considered quantizations, but of all Fock quantizations which, respecting unitary evolution and spatial symmetry invariance, are based on variables which satisfy dynamical equations of the KG type with a mass that varies in time --even if one allows for mass corrections that vanish 
asymptotically in the ultraviolet sector-- will be the subject of future research. Besides, the same kind of quantization procedure can be applied to tensor perturbations of a flat FRW spacetime with compact spatial sections. These tensor perturbations describe cosmological gravitational waves which, after a proper scaling and in conformal time, satisfy a wave equation of the KG type with time dependent mass \cite{bar}. Hence, our analysis can be applied to these tensor perturbations as well.

Our results are also useful in the context of the hybrid quantization of inhomogeneous cosmological systems \cite{gowdy}, where the Fock description of the inhomogeneities is combined with a polymeric quantization of the homogeneous background geometry, as predicated by loop quantum cosmology (LQC) \cite{LQCap,lqc}. This hybrid quantization was applied for the first time and developed in the case of vacuum Gowdy cosmologies with spatial sections of three-torus topology \cite{gowdy,gowdy1}, and it was later extended by including matter \cite{gowdy_matt}. Recently, this hybrid quantization has been applied to the complete quantization of the inflationary system mentioned above, though in the case where the spatial topology is that of a three-sphere \cite{hybrid-pert}. The unique privileged Fock quantization selected by our criteria (including both the scaling of the field and the representation of the CCR's) is used for the quantum description of the scalar perturbations, while the FRW geometry (with positive
spatial curvature), together with a massive homogeneous scalar field, is quantized by employing the techniques of LQC. A similar procedure can be followed in the more realistic case of three-torus topology, where the background corresponds to a flat FRW spacetime. For this situation, the results presented here can be used to obtain a unique Fock quantization of the scalar perturbations, assuming that the system admits a regime in which the deparametrized description of the perturbations is valid. Notice that this regime will cover only a restricted time interval; hence the importance that our results do not depend on the specific time domain under consideration (and in particular on whether it is bounded or not).

In conclusion, we have presented a set of criteria to select a preferred unique class of Fock quantizations for scalar fields in a variety of nonstationary spacetimes with compact spatial sections of three-torus topology. This includes cosmological systems which provide a large scale description of the Universe (at least as far as scalar cosmological perturbations are considered). The application of our results to the quantization of scalar fields in such realistic scenarios allows one to extract robust physical predictions from the quantum realm, which might be confronted with observations.

\acknowledgments{The authors are grateful to S. Carneiro, M. Fern\'andez-M\'endez, M. Mart\'{\i}n-Benito, J. Olmedo, and T. Pereira for discussions. This work was
supported by the research grants MICINN/MINECO FIS2011-30145-C03-02 from Spain, DGAPA-UNAM IN117012-3 from Mexico, and
CERN/FP/116373/2010 from Portugal. D. M-dB was
supported by CSIC and the European Social Fund under the grant JAEPre\_09\_01796.}

\appendix

\section{Invariant complex structures: Complex modes}

In this appendix we will consider the Fourier decomposition of the field \eqref{cfd1} in terms of complex eigenfunctions of the LB operator and
determine again the form of the invariant complex structures, but this time in the corresponding basis of complex modes. Obviously, the result will reproduce our previous conclusion,
but the derivation may be helpful to clarify certain aspects of the role played by the irreducible representations of the symmetry group when these are complex.

The action on the field of $T_{\vec{\alpha}}$ leads to the following (active) transformations of the complex modes:
\begin{equation}
\label{cmt1}
\mathfrak{q}'_{\vec{m}}=e^{-i\vec{m}\cdot\vec{\alpha}}\mathfrak{q}_{\vec{m}}, \qquad
\mathfrak{p}'_{\vec{m}}=e^{-i\vec{m}\cdot\vec{\alpha}}\mathfrak{p}_{\vec{m}}.
\end{equation}
Hence, every transformation $T_{\vec{\alpha}}$ acts on both the complex field modes $\{\mathfrak{q}_{\vec{m}}\}$ and the complex momentum modes $\{\mathfrak{p}_{\vec{m}}\}$ as the product of three copies of $U(1)$, one for each direction. It is also clear that, for each label $\vec{m}$, we get a different irreducible representation of this group. At this point, we can make use of the Schur's lemma \cite{kirill}, which tells us that every symmetry invariant map on phase space must be block diagonal [in the basis of complex modes $\{(\mathfrak{q}_{\vec{m}},\mathfrak{p}_{\vec{m}})\}$],
with $2\times2$ blocks given by arbitrary complex matrices. Nonetheless, since $J$ is a real map on phase space, so that it preserves the reality conditions on the modes, the
blocks $J_{\vec{m}}$ and $J_{-\vec{m}}$ must be related by complex conjugation. Then, reordering our complex basis in the form
$\{(\mathfrak{q}_{\vec{m}},\mathfrak{q}_{-\vec{m}},\mathfrak{p}_{-\vec{m}},\mathfrak{p}_{\vec{m}})\}$ (with $m$ such that it has at most one negative component),
we conclude that the $4\times4$ blocks of $J$ must have the expression
\begin{equation}
J_{\vec{m}}=\left(\begin{array}{cc}
J^{\mathfrak{qq}}_{\vec{m}} &  J^{\mathfrak{qp}}_{\vec{m}}\\
J^{\mathfrak{pq}}_{\vec{m}} &  J^{\mathfrak{pp}}_{\vec{m}}
\end{array}
\right), \qquad
J_{\vec{m}}^{\mathfrak{rs}}=\left(\begin{array}{cc}
\gamma_{\vec{m}}^{\mathfrak{rs}} & 0\\
0 &  (\gamma_{\vec{m}}^{\mathfrak{rs}})^{\ast}
\end{array}
\right),
\end{equation}
where the superindex $\mathfrak{rs}$ stands for any of the four possible sub-blocks ($\mathfrak{qq}$,  $\mathfrak{qp}$, $\mathfrak{pq}$, or $\mathfrak{pp}$), and $\gamma_{\vec{m}}^\mathfrak{rs}$ is a complex number.

We still have to impose the conditions on $J$ ensuring that it is a compatible complex
structure. We first require that $J_{\vec{m}}^{T}\Omega_{\vec{m}}$ be a Hermitian positive definite matrix,
where $\Omega_{\vec{m}}$ has the same form as $\Omega_{\vec{n}_j}$, given in Eq. \eqref{njss}.
It then follows that the matrices
$J^{\mathfrak{pq}}_{\vec{m}} $ and $-J^{\mathfrak{qp}}_{\vec{m}} $ must be positive definite
and that $J^{\mathfrak{qq}}_{\vec{m}} =(J^{\mathfrak{pp}}_{\vec{m}} )^{\ast}$. Therefore,
$\gamma^{\mathfrak{pq}}_{\vec{m}} =d^{(1)}_{\vec{m}}$ and $-\gamma^{\mathfrak{qp}}_{\vec{m}} =d^{(2)}_{\vec{m}}$ must be positive
numbers, and $\gamma^{\mathfrak{qq}}_{\vec{m}}=(\gamma^{\mathfrak{pp}}_{\vec{m}})^{\ast}$.
Moreover, the conditions
$J_{\vec{m}}^{T}\Omega_{\vec{m}}J_{\vec{m}}=\Omega_{\vec{m}}$ (namely, that $J$ be a symplectomorphism) and
$J_{\vec{m}}^{2}=-\mathbf{I}_{4\times4}$ (i.e., that $J^2$ be minus the identity) imply then that $\gamma^{\mathfrak{qq}}_{\vec{m}}=
\pm \sqrt{d^{(1)}_{\vec{m}}d^{(2)}_{\vec{m}}-1}$. In conclusion, every invariant complex structure
$J$ is block diagonal, with $4\times4$ blocks that, in the considered basis of complex modes, take the form
\begin{equation}
J_{\vec{m}}=\left(\begin{array}{cc}
\pm d^{(3)}_{\vec{m}}\,\mathbf{I}_{2\times2} & -d^{(2)}_{\vec{m}}\, \mathbf{I}_{2\times2} \\
d^{(1)}_{\vec{m}}\,\mathbf{I}_{2\times2} & \mp d^{(3)}_{\vec{m}}\,\mathbf{I}_{2\times2}
\end{array}\right),
\end{equation}
where the $d_{\vec{m}}$'s are positive numbers
satisfying $d^{(1)}_{\vec{m}}d^{(2)}_{\vec{m}} \ge 1$ and $d^{(3)}_{\vec{m}}=\sqrt{d^{(1)}_{\vec{m}}d^{(2)}_{\vec{m}}-1}$.

Of course, the two forms obtained for the invariant complex structures, using either real or complex eigenfunctions of the LB operator, are equivalent. They are related by a change of
basis in phase space, passing from the real to the complex modes:
\begin{equation}
 q_{\vec{n}_j}=\frac{1}{2^{3/2}}\left.\left(\mathfrak{q}_{\vec{m}}+\mathfrak{q}_{-\vec{m}}\right)\right|_{\vec{m}=\vec{n}_{j}},
\qquad x_{\vec{n}_j}=\frac{i}{2^{3/2}}\left.\left(\mathfrak{q}_{\vec{m}}-\mathfrak{q}_{-\vec{m}}\right)\right|_{\vec{m}=\vec{n}_{j}},
\end{equation}
together with a similar change from $\{(p_{\vec{n}_j}, y_{\vec{n}_j})\}$ to
$\{(\mathfrak{p}_{\vec{m}}, \mathfrak{p}_{-\vec{m}})\}$. It is then straightforward to check
that, in fact, $c_{\vec{n}_j}^{(i)}=d^{(i)}_{\vec{m}}$ for $i=1,2,3$.

\end{document}